\documentclass[journal]{IEEEtran}
\usepackage{comment}
\usepackage{algorithm}
\usepackage{array}
\usepackage{url}
\usepackage{verbatim}
\usepackage{tikz}
\usepackage{circuitikz}
\usetikzlibrary{positioning}
\usepackage{dsfont}
\usepackage{hyperref}
\usepackage[caption=false,font=normalsize,labelfont=sf,textfont=sf]{subfig}
\usepackage{stfloats}
\usepackage{multirow}
\usepackage{cite}
\usepackage{amsmath,amssymb,amsfonts}
\usepackage{algorithmic}
\usepackage{graphicx}
\usepackage{textcomp}
\usepackage{xcolor}
\usepackage{fancybox}
\def\BibTeX{{\rm B\kern-.05em{\sc i\kern-.025em b}\kern-.08em
    T\kern-.1667em\lower.7ex\hbox{E}\kern-.125emX}}
\usepackage{balance}    
\usetikzlibrary{shapes, arrows}

\begin{document}

\title{TSO–DSO Coordinated Reactive Power Dispatch for Smart Inverters with Multiple Control Modes — Real-Time Implementation }

\author{Mohammad~Almomani,~\IEEEmembership{graduate student member,~IEEE,}
        Ahmed~Alkhonain,~\IEEEmembership{graduate student member,~IEEE,}
        and~Venkataramana~Ajjarapu,~\IEEEmembership{Fellow,~IEEE}


\IEEEauthorblockA{\textit{Department of Electrical \& Computer Engineering, Iowa State University, Ames, IA, USA} \\
Emails: \{mmomani, ahmedkh, vajjarap\}@iastate.edu}}

\maketitle
\begin{abstract}
This paper presents TSO–DSO coordinated reactive power dispatch, with a focus on real-time implementation. A sensitivity-aware, mixed-integer linear programming (MILP) formulation is developed to model the IEEE 1547-compliant droop-based control modes—Volt-VAR (VV), Volt-Watt (VW), and Watt-VAR (WV)—of smart inverters. The algorithm employs a hierarchical optimization strategy using Special Ordered Sets (SOS1) to enhance computational efficiency and supports limited measurement scenarios through Recursive Least Squares (RLS) estimation. The proposed method is tested on the IEEE 13-bus and 123-bus distribution networks, which are connected to a 9-bus transmission system. Results demonstrate the feasibility and effectiveness of the real-time dispatch framework in improving voltage regulation and minimizing power curtailment. 

\doublebox{\parbox{0.8\linewidth}{
\centering
This paper has been submitted to IEEE Transactions on Smart Grids.
}}

\end{abstract}
\section{Introduction}

The rapid growth of Distributed Energy Resources (DERs)—including photovoltaic (PV) systems, wind turbines, and battery energy storage systems (BESSs)—has transformed modern power system operations. Traditionally, voltage regulation and reactive power support were managed by traditional resources at the transmission level. However, the increasing share of DERs connected at the distribution level now requires coordinated voltage support between Transmission System Operators (TSOs) and Distribution System Operators (DSOs) to ensure stability across normal and disturbed conditions~\cite{IEEE1547,FERC2222}. 

To facilitate this transition, the IEEE 1547-2018 standard provides guidelines for the interconnection and interoperability
of DERs with the electric grid. It specifies five operational modes for voltage regulation: two constant modes (Power Factor and Reactive Power) and three droop-based modes (Volt–VAr, Volt–Watt, and Watt–VAr). Sections~4.6.3 and~5.3.1 specify that DERs must transition between modes within 30~s of receiving commands from the Area EPS operator~\cite{IEEE1547}. 

Among DER technologies, photovoltaic systems have experienced the most rapid growth over the past decade. However, their widespread deployment introduces several challenges for distribution networks, including voltage rise, reverse power flow, and fluctuations, particularly
during periods of high solar generation and low demand \cite{Ref2}. These problems are further exacerbated by the limited hosting capacity of the distribution feeders. As a result, voltage control has become a critical function of the Distribution Management System (DMS), which is tasked with maintaining service voltages within regulatory limits, such as those defined by the American National Standards Institute (ANSI).

Smart inverters (SIs)—compliant with IEEE~1547 and regional standards such as California Rule~21~\cite{Ref4} and Hawaii Rule~14H~\cite{Ref5}—enable autonomous voltage and frequency support through local control functions~\cite{Ref6,Ref7}.  Unlike traditional inverters that disconnect during voltage
deviations, smart inverters are capable of continuous operation under abnormal conditions due to their self-regulating and adaptive features \cite{Ref6}. Their capabilities are essential to mitigate voltage rise, reverse power flow, and overvoltage problems in high-DER feeders~\cite{Ref2}, increasing the host capacity of distribution feeders, but their heterogeneous responses make centralized coordination challenging~\cite{Sen2,Der1,Der}. Specifically, the combination of Volt-Var (Q(V)), watt-var (Q(P)), and Volt-Watt (P(V)) functionalities is particularly beneficial for
overvoltage mitigation and maximizing PV hosting capacity. As the share of inverter-based DERs continues to rise, their integration into distribution grid operations becomes increasingly critical. However, despite the availability of standardized control modes, the
heterogeneity of inverter technologies and their associated control strategies complicates their incorporation into centralized voltage
support and reactive power dispatch frameworks \cite{Sen2}, \cite{Der1}. Moreover, conventional reactive power dispatch methods—designed for centralized synchronous generators—do not account for the decentralized, autonomous control behavior of smart inverters \cite{Der}.

Despite extensive work on reactive power dispatch and Distribution Optimal Power Flow (DOPF) formulations~\cite{Ref9,Ref10,Ref11}, most existing approaches neglect IEEE~1547-mandated droop-based responses.  As a result, control setpoints derived from DOPF may violate local inverter operating rules, rendering such methods ineffective at the device level. Recent efforts have incorporated simplified or linearized droop representations~\cite{Ref8,Ref14,conf_paper,Sen}, or aggregated capability curve estimation~\cite{Der,Der3}, but these often assume single-phase or balanced systems, overlooking the coupling effects and nonlinearity inherent in multiphase networks~\cite{ref16,Ref17}. Moreover, while convex and MILP relaxations improve scalability~\cite{Ref13}, comprehensive optimization of both droop settings and inverter modes within real-time TSO–DSO frameworks remains largely unexplored.

 In our previous research, the Sensitivity-Aware Reactive Power Dispatch method introduced in \cite{Sen}, which prioritizes DERs based on their sensitivity to substation voltage. This approach effectively reduces the number of required control signals, making it especially suitable for large-scale networks where communication overhead is a concern. While this method is efficient for systems dominated by PQ-mode inverters, it does not fully capture the local operational modes of smart inverters, such as VV, VW, and WV modes. To enhance the capability of dispatch algorithms, the authors incorporated a simplified VV droop function into the dispatch logic in \cite{conf_paper}, treating it as a linear Q(V) characteristic. Although this addition improves the representation of smart inverter behavior, it does not address the piecewise, nonconvex, nonlinear nature of standard droop curves defined in IEEE 1547. Further advancing the understanding of DERs’ capabilities, in \cite{Der} we introduced a method to derive the aggregated reactive power capability curve at the TSO-DSO interface. This technique provides TSOs with a reliable estimate of the reactive power support available from DERs. The resulting VAR-capability curve facilitates coordinated dispatch decisions across multiple DERs, improving system reliability during contingencies \cite{Der3}.

Motivated by these gaps, this work proposes a real-time, IEEE~1547-compliant {Transmission–Distribution Coordinated Dispatch Algorithm (TDCDA)} that jointly optimizes operation modes and droop parameters for smart inverters. The framework enables standards-compliant and efficient DER coordination across unbalanced multiphase distribution systems with limited measurement capabilities. This work fills the identified gaps in coordinated reactive power dispatch by jointly optimizing smart inverter \emph{operation modes} and \emph{droop parameters} under IEEE~1547 requirements. The main contributions are summarized as follows:

\begin{enumerate}
    \item \textbf{Unified MILP Formulation of Droop-Based Inverter Controls:}  
    A mixed-integer linear programming (MILP) formulation is developed to model piecewise-linear \textbf{Volt–VAr}, \textbf{Volt–Watt}, and \textbf{Watt–VAr} droop functions using the Big-M technique~\cite{Big_M}. These models are embedded into the LineDistOPF~\cite{P19} framework, allowing simultaneous optimization of inverter modes and droop curves within a single coordinated dispatch problem.

    \item \textbf{Real-Time Feasibility via Hierarchical and Observable Reformulation:}  
    To ensure real-time applicability, the LineDistOPF is reformulated to distinguish between observable and unobservable states, accommodating limited field measurements. A hierarchical optimization strategy enables efficient mode and segment selection, significantly reducing computational time.
\end{enumerate}
 {Integrated TDCDA for Multiphase Systems with Droop Mode Coordination:}  
    Building upon prior works~\cite{Der,Sen,conf_paper}, the proposed real-time TSO–DSO coordination framework integrates IEEE 1547-compliant droop-based control functions. Unlike \cite {Der}, which derived day-ahead capability curves without considering local controller behavior, this approach determines real-time capability while explicitly considering inverter behavior. Compared with~\cite{conf_paper} and~\cite{Sen}, it extends sensitivity-based dispatch and simplified linear droop modeling to realistic nonlinear characteristics suitable for online operation. The proposed algorithm optimizes both {control modes and droop settings} in real time, ensuring practicality and compliance with IEEE~1547.

\section{Mathematical Models}
\begin{figure}
    \centering
    \includegraphics[width=0.5\textwidth]{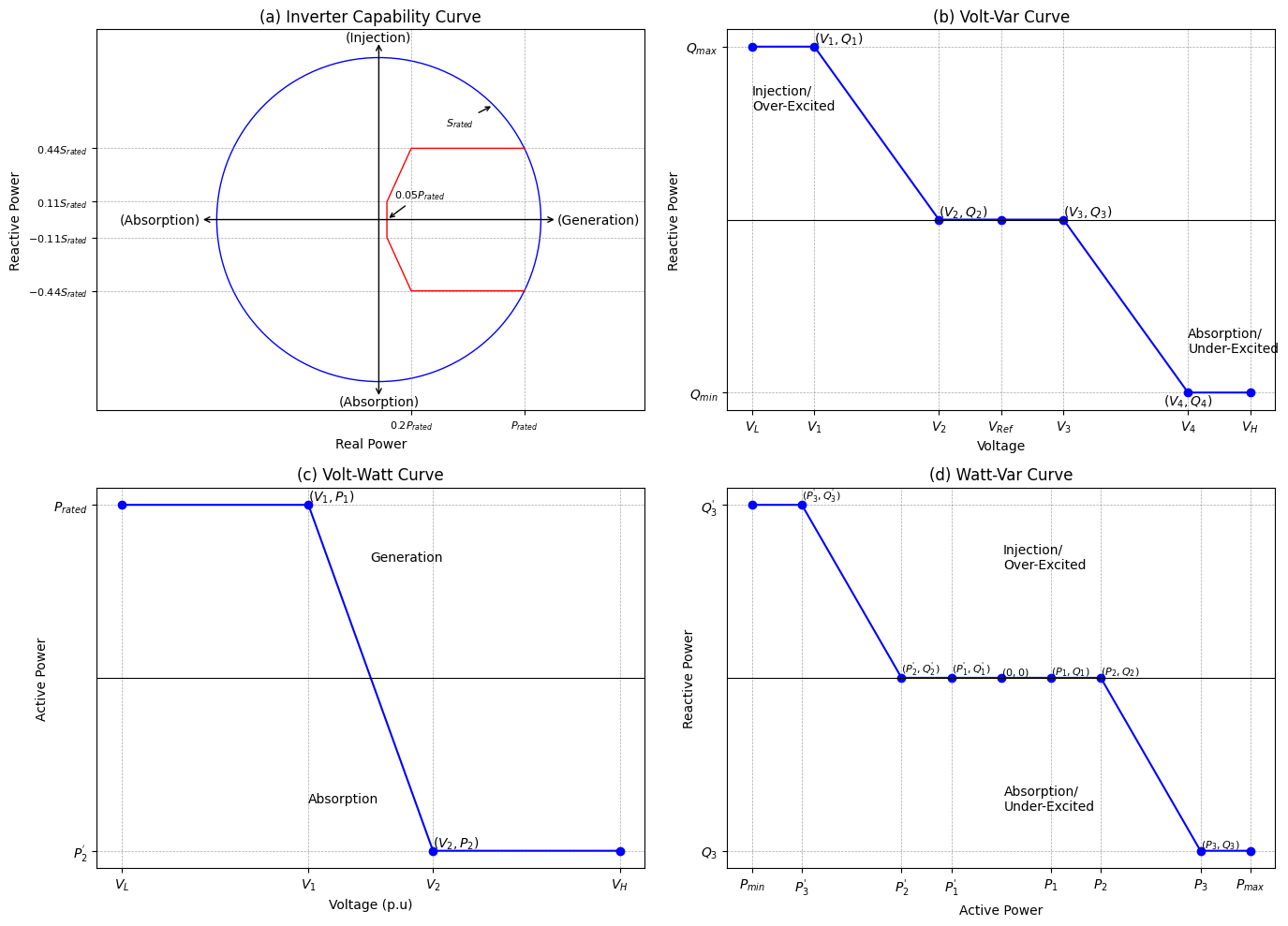}
    \caption{
        Inverter control characteristics \cite{IEEE1547}. 
        (a) The inverter P-Q capability curve.
        (b) The Volt-VAR curve.
        (c) The Volt-Watt curve.
        (d) The Watt-VAR curve.
            }
\label{fig:operating_regions}
\end{figure}

Mathematical models of the Inverter capability curve and the smart inverter control modes are established based on the IEEE 1547 standard, as shown in Figure \ref{fig:operating_regions}.
\subsection{Inverter Capability Characteristic}
According to IEEE 1547 \cite{IEEE1547}, the operating regions of a PV inverter are determined by both real (\(P\)) and reactive (\(Q\)) power limits, constrained by the inverter's apparent power rating (\(S_{\text{rated}}\)). These regions are governed by three key constraints: Linear, Maximum, and minimum limits and nonlinear constraints, as illustrated in Fig.~\ref{fig:operating_regions} (a). The nonlinear constraints \(P^2 + Q^2 \leq S_{\text{rated}}^2\) can be addressed by linearizing them using an 8-vertex polygon approximation \cite{Vertex_Polygon}. So, the inverter capability can be modeled as: 
    \begin{subequations}
        \begin{align}
           - m_{PQ} P^G - b_{PQ} & \leq Q^G \leq m_{PQ} P^G + b_{PQ}
\\        
  Q_{\text{min}} &  \leq Q^G \leq Q_{\text{max}}\\
 P_{\text{min}}  & \leq P^G \leq P_{\text{max}}.\\
    -S_{rated} & \leq\cos( \gamma_l) P + \sin( \gamma_l) Q \leq s_{rated} 
        \end{align}
    \label{eq:cap}
    \end{subequations}
where 
\(
\gamma_l = \left( \frac{2l}{7}-1\right) asin(\frac{Q_{max}}{S_{rated}}), \quad l = 0 \dots 7, \quad  \forall  \mathbf{G}^{PV}
\)
 The default values of $m_{PQ}, b_{PQ}$ from IEEE 1547 are 2.2 and 0, respectively. These constraints ensure the inverter operates within safe and defined limits while maintaining flexibility for reactive power injection or absorption, as required by system conditions.

\subsection{Smart Inverter Droop Functions}

Smart inverters enable autonomous control by implementing droop-based control functions that establish relationships between power output and local measurements. The three predominant control modes defined for smart inverters are Volt-VAR (\(Q(V)\)), Volt-Watt (\(P(V)\)), and Watt-VAR (\(Q(P)\)). These modes are formalized in the IEEE 1547-2018 standard~\cite{IEEE1547}, each characterized by a piecewise-linear control law that governs inverter response as a function of measured electrical quantities. To incorporate these control characteristics into an optimization framework, the \textbf{big-M method} is employed, which facilitates the modeling of piecewise-linear relationships through linear constraints and binary variables.

The Volt-VAR mode (\(Q(V)\)), illustrated in Fig.~\ref{fig:operating_regions}(b), defines a continuous, univariate, piecewise-linear mapping between reactive power output and voltage magnitude at the point of common coupling (PCC)~\cite{IEEE1547}. Similarly, the Volt-Watt mode (\(P(V)\)) specifies how the inverter’s active power output is modulated based on the voltage at the PCC, with the control curve—shown in Fig.~\ref{fig:operating_regions}(c)—designed to curtail active power as voltage exceeds a specified threshold, thereby mitigating overvoltage conditions. Lastly, the Watt-VAR mode (\(Q(P)\)) establishes a relationship between reactive power (\(Q\)) and active power (\(P\)) output of the inverter, enabling dynamic reactive power support (either injection or absorption) based on real-time active power conditions.

The mathematical formulation of the Volt-VAR droop curve, denoted as \( Q(V) \), is expressed as a piecewise-linear function. 
To incorporate this control logic into a mixed-integer linear programming (MILP) formulation, binary variables \( z_{l}^{VV} \in \{0,1\} \) are introduced to indicate which of the five segments is active. using the big-M method, the piecewise-linear function can be reformulated as: 
\begin{subequations}
\begin{align}
&(z_{l}^{VV} - 1)M + V_{l,\text{min}} \leq V^G \leq V_{l,\text{max}} + (1 - z_{l}^{VV})M\label{eq:vv1} \\
&Q^G \geq m_l^{VV} V^G + b_l^{VV} - M(1 - z_{l}^{VV}) \label{eq:vv2a} \\
&Q^G \leq m_l^{VV} V^G + b_l^{VV} + M(1 - z_{l}^{VV}) \label{eq:vv2b}\\
&\sum_{l=1}^{5} z_{l}^{VV} = 1, \quad z_{l}^{VV} \in \{0, 1\}, \quad \forall l \in [1,2,3,4,5]\label{eq:vv3}
\end{align}
\label{eq:vv}
\end{subequations}
Where M is a big number an equation \eqref{eq:vv1} ensures that the generator terminal voltage \( V^G \) lies within the bounds of the active segment \( l \). Equations \eqref{eq:vv2a} and \eqref{eq:vv2b} ensure that the computed reactive power \( Q^G \) lies on the line defined by the slope \( m_l^{VV} \) and intercept \( b_l^{VV} \) of the active segment which are found from the curve in Figure \ref{fig:operating_regions}. To ensure that only one segment is active at any given time, the exclusivity condition is enforced in Equation \eqref{eq:vv3}.

A similar modeling structure is used for the Volt-Watt (VW) control mode. Binary variables \( z_{l}^{VW} \in \{0,1\} \), where \( l \in \{1, 2, 3\} \), denote the active segment. The Model is given by: 
\begin{subequations}
\begin{align}
&(z_{l}^{VW} - 1)M + V_{l,\text{min}} \leq V^G \leq V_{l,\text{max}} + (1 - z_{l}^{VW})M
\label{eq:vw1}\\
&P^G \geq m_l^{VW} V^G + b_l^{VW} - M(1 - z_{l}^{VW}) \label{eq:vw2a}\\
&P^G \leq m_l^{VW} V^G + b_l^{VW} + M(1 - z_{l}^{VW}) \label{eq:vw2b}\\
&\sum_{l=1}^{3} z_{l}^{VW} = 1, \quad z_{l}^{VW} \in \{0, 1\}, \quad , \quad \forall l \in [1,2,3]
\label{eq:vw3}
\end{align}
\label{eq:vw}
\end{subequations}
The Watt-VAR (WV) control mode is modeled similarly, with binary variables \( z_{l}^{WV} \in \{0,1\} \), where \( l \in \{1, \ldots, 5\} \):
\begin{subequations}
\begin{align}
&(z_{l}^{WV} - 1)M + P_{l,\text{min}} \leq P^G \leq P_{l,\text{max}} + (1 - z_{l}^{WV})M
\label{eq:wv1}\\
&Q^G \geq m_l^{WV} P^G + b_l^{WV} - M(1 - z_{l}^{WV}), \quad \forall l \label{eq:wv2a}\\
&Q^G \leq m_l^{WV} P^G + b_l^{WV} + M(1 - z_{l}^{WV}), \quad \forall l
\label{eq:wv2b} \\
&\sum_{l=1}^{5} z_{l}^{WV} = 1, \quad z_{l}^{WV} \in \{0, 1\}, \quad \forall l \in [1,2,3,4,5]\label{eq:wv3}
\end{align}
\label{eq:wv}
\end{subequations}
This unified MILP-based formulation enables systematic incorporation of smart inverter droop-based control logic into power system optimization problems.

\subsection{Distribution Network Model}

Distribution networks are inherently unbalanced due to the presence of untransposed lines, uneven load distributions, and multiphase connections (see [\cite{P25}, Ch. 2]). This necessitates the inclusion of coupling effects between phases in their modeling. To capture these dynamics, a radial unbalanced distribution network is represented as a directed graph comprising a set of buses \( \mathcal{B} \) and branches \( \mathcal{L} \) connecting these buses.

In a three-phase system, the power flow is described using the branch flow model (BFM) equations \cite{P25}. 

The equations 
represent a nonlinear formulation that accurately models the system by capturing mutual coupling between phases. However, its nonlinear nature makes it computationally challenging for optimization. To simplify the BFM, the linearized \textit{LinDist3Flow} formulation is adopted, which linearizes the nonlinear equations while preserving the coupling between phases. Following \cite{P19}, the voltage phasor ratios are approximated by:
\(
V_i^b (V_i^a)^{-1} \approx \alpha, \quad V_i^c (V_i^a)^{-1} \approx \alpha^2,
\)
where:
\(
\alpha = e^{j 120^\circ} \)
By linearizing the BFM equations and ignoring loss terms, 
the voltage difference between two nodes \( i \) and \( j \) is related to the power flows and line impedance as follows:
\begin{equation}
V_i^* V_i - V_j^* V_j = -Z_{ij}^P P_{j} - Z_{ij}^Q Q_{j}, \quad \forall (i, j) \in \mathcal{L}.
\label{eq:DN3}    
\end{equation}

Where:  \( V_j = [V_{a,j}, V_{b,j}, V_{c,j}]^T \): voltage phasors at node \( j \) for phases \( a, b, \) and \( c \). \( P_{j} = [P_{a,j}, P_{b,j}, P_{c,j}]^T \) and \( Q_{j} = [Q_{a,j}, Q_{b,j}, Q_{c,j}]^T \): active and reactive power injections at node \( j \), respectively. \( Z_{ij}^P \) and \( Z_{ij}^Q \): matrices representing the line impedance between nodes \( i \) and \( j \) for active and reactive power flows, considering mutual coupling between phases, expressed as \cite{PP25}. The nodal voltages can be expressed in terms of a matrix representation. Define the squared voltage magnitudes \( Y_j = [|V_{a,j}|^2, |V_{b,j}|^2, |V_{c,j}|^2]^T \) and reference the substation voltage as \( Y_0 \). The compact representation of nodal voltages becomes:
\begin{equation}
\begin{bmatrix}
M_0 & M^T
\end{bmatrix}
\begin{bmatrix}
Y_0 \\
Y
\end{bmatrix} = Z_D^P P + Z_D^Q Q
\label{eq:DN4}
\end{equation}
where \( M_0 \) is a matrix mapping the substation, and \( M \) corresponds to the network connectivity. $ Z_D^P,  Z_D^Q$ are the system impedance matrices.

\subsection{First-Order Linearization of Voltage Magnitude}

The droop characteristic typically establishes the relationship between voltage magnitude and active/reactive power \(P,Q \sim m_l V\). In the context of LinDistFlow equations, this relationship is inherently nonlinear, as the equations are linear with respect to the square of the voltage magnitude, denoted as \( Y = V^2 \), and the power. Consequently, when representing the droop characteristic in terms of \( Y \), the relationship becomes nonlinear.
To address this nonlinearity, we can decompose \( Y - Y_0 \), where \( Y_0 \) represents the square of the substation voltage magnitude, as:
\(
Y - Y_0 = (V - V_0)(V + V_0),
\)
where \( V_0 \) is the substation voltage magnitude. In this expression, the dominant coefficient is \( (V - V_0) \), particularly when the voltage \( V \) is close to its nominal value of 1 p.u. By assuming that \( V + V_0 \approx 2 V_0 \) (since the voltage magnitude near the nominal condition implies this approximation), the droop characteristic can be linearized in terms of \( Y \) and power. The resulting linearized droop characteristic becomes:
\begin{equation}
 V=\frac{Y }{2 \sqrt{Y_0}} +\frac{\sqrt{Y_0}}{2}
\label{eq:FO1}    
\end{equation}
 
To account for the voltage dependency of loads, the power injection at each node is modeled as:
\begin{subequations}
    \begin{align}
P_{j}^{L} = P_{j}^{L,0} (a_0 + a_1 |V_j|^2+a_2 |V_j|) \\
Q_{j}^{L} = Q_{j}^{L,0} (a_0 + a_1 |V_j|^2+a_2 |V_j|)
    \end{align}
    \label{eq:DN5}
\end{subequations}
where \( a_0, a_1 \) and \( a_2 \) are coefficients representing constant power, constant impedance, and constant current loads, respectively. Similarly, the power-voltage relationship is linear for a constant current load. To ensure a linear relationship between the squared voltage and power for this type of load, the same First-Order Linearization of voltage magnitude in Equation \eqref{eq:FO1} is applied. Consequently, Equation \ref{eq:DN5} can be reformulated as:
\begin{subequations}
    \begin{align}
P_{j}^{L} = P_{j}^{L,0} \left[a_0 + a_1 Y_j+a_2 \left(\frac{Y_j }{2 \sqrt{Y_0}} +\frac{\sqrt{Y_0}}{2}\right)\right] \\
Q_{j}^{L} = Q_{j}^{L,0} \left[a_0 + a_1 Y_j+a_2 \left(\frac{Y_j }{2 \sqrt{Y_0}} +\frac{\sqrt{Y_0}}{2}\right)\right]
    \end{align}
    \label{eq:FO2}
\end{subequations}
The nodal voltages \( Y \) can then be computed from equation (\ref{eq:DN4}) as:
\begin{align}
KY &= Y_0 + R^{eq} P^G + X^{eq} Q^G \nonumber \\
   & - \left( R^{eq} \,\mathbb{D}(P^L) + X^{eq} \, \mathbb{D}(Q^L) \right) 
   \left( A_0 + A_2 \frac{\sqrt{Y_0}}{2} \right)
\label{eq:FO3}
\end{align}

where:
\(
R^{eq} = -M^T Z^P M^{-1}, \quad X^{eq} = -M^T Z^Q M^{-1},
\) and \( K = I +R^{eq} \, \mathbb{D} \left( \mathbb{D}(P^L) \left( A_1 + \frac{A_2}{2\sqrt{Y_0}} \right) \right)  +\; X^{eq} \, \mathbb{D} \left( \mathbb{D}(Q^L) \left( A_1 + \frac{A_2}{2\sqrt{Y_0}} \right) \right) 
\)
Noting that, because of the radial network $(-M^{-T}M_0Y_0)$
is same as $Y_0 \mathbf{1}$ and M and K both are invertible matrices. \(A_0, A_1, A_2\) are vectors of \(a_0, a_1, a_2\) elements, respectively. \(\mathbb{D}(\cdot)\) is a diagonal operator.  Finaly, the line flow equation (from BFM model) can be rewritten as follows, assuming negligible losses:
\begin{subequations}
    \begin{align}
M P_{tl} = P^G & -  \mathbb{D}(P^L) \left(  A_0+A_2 \frac{\sqrt{Y_0}}{2} \right) \notag\\&- \mathbb{D}(P^L)  \mathbb{D}\left(A_1+  \frac{A_2 }{2\sqrt{Y_0}}\right) Y \\
M Q_{tl} = P^G - & \mathbb{D}(Q^L) \left(  a^0+a_2 \frac{\sqrt{Y_0}}{2} \right)\notag\\&- \mathbb{D}(Q^L) \mathbb{D} \left( A_1+  \frac{A_2 }{2\sqrt{Y_0}}\right) Y
    \end{align}
    \label{eq:FO4}
\end{subequations}

\section{Problem Formulation for Real-Time Application}



The integration of three operational modes—Volt-VAR, Volt-Watt, and Watt-VAR—provides flexibility to dynamically adjust the inverter's behavior in response to changing grid conditions. According to IEEE 1547 \cite{IEEE1547} Section 4.6.3, the transition between operational modes must begin within 30 seconds of receiving the mode setting change at the local DER communication interface and be completed, following a smooth transition, within 300 seconds. To achieve optimal operation, these modes must be integrated into a unified optimization framework, enabling the system to select the most suitable mode at each node. This can be formulated based on the previous formulation by changing the segment exclusion constraints as follow: 
\begin{equation}
\sum_{l=1}^5 z_{l}^{VV}+\sum_{l=1}^3 z_{l}^{VW}+ \sum_{l=1}^5 z_{l}^{WV} =1, \quad  z_{l}^{WV},z_{l}^{VV},z_{l}^{VW} \in \{0, 1\}
\label{eq:MILP}
\end{equation}

\begin{table}
\centering
\caption{Summary of inverter control variables.}
\resizebox{\linewidth}{!}{%
\begin{tabular}{|c|c|c|c|}
\hline
\textbf{Control Mode} & \textbf{Binary Variable} & \textbf{Setting variable} & \textbf{Operational variable} \\
\hline
Volt-VAR  & $z_l^{VV}$ & $V_{\text{set}}^{VV} = V_3 - V_{\text{ref}} $ & $P^G$ \\
\hline
Volt-Watt & $z_l^{VW}$ & $V_{\text{set}}^{VW} = V_1$ & $Q^G$ \\
\hline
Watt-VAR  & $z_l^{WV}$ & $P_{\text{set}}^{WV} = P_2 $ & $P^G$ \\
\hline
\end{tabular}
}
\label{tab:Table1}
\end{table}
Table \ref{tab:Table1} shows the optimization variables: binary and continuous (setting variables and operation variables) of each control mode. The optimization framework targets the \emph{offset control variables} ($b_l$), while maintaining constant \emph{droop slopes} ($m_l$) as recommended by IEEE 1547 to preserve dynamic stability. To further reduce the complexity of the control space, symmetry around the y-axis is assumed in each droop curve, thereby minimizing the number of independent variables per mode. The following explanation outlines the setting variables used to optimize the droop curves for each control mode. 

\begin{itemize}
    \item \textbf{Volt-VAR:} The setting variable is defined as \( V_{\text{set}}^{VV} = V_3 - V_{\text{ref}} = V_{\text{ref}} - V_2 \), with the width \( V_4 - V_3 \) fixed. The real power \( P^G \) is used as the operating point.
    \item \textbf{Volt-Watt:} The setting variable is \( V_{\text{set}}^{VW} = V_1 \), assuming \( V_2 - V_1 \) remains constant. The inverter controls the reactive power \( Q^G \) as the operating point.
    \item \textbf{Watt-VAR:} The setting variable is \( P_{\text{set}}^{WV} = P_2 = -P_2' \), while \( P_3 - P_2 \) and \( P_2' - P_3' \) are held at default values. 
\end{itemize}

This formulation ensures that only the offsets \( b_l^{VV} \), \( b_l^{VW} \), and \( b_l^{WV} \) are optimized, preserving linearity in the optimization model.

\subsection{Hierarchical Optimization Approach for Operational Modes and Segments}

As shown in Table~\ref{tab:Table1}, each DER requires 13 binary variables for control mode and segment selection, along with six continuous variables for offset and operational settings. For a system with \( n \) DERs, this results in \( 13 \times n \) binary variables, posing significant computational challenges in real-time applications. To address this, we implement a mode exclusivity constraint using special order set type 1 (SOS1) \cite{SOS}  and propose a hierarchical solution strategy to enhance scalability and tractability.
 Each DER operates in one of three modes—Volt-VAR, Volt-Watt, or Watt-VAR—where only one mode and a single segment within that mode can be active at any time.  SOS1 allows only one variable in a defined group to be non-zero, without the need for additional binary variables. Solvers like Gurobi and CPLEX exploit this property natively, resulting in faster convergence and a reduced search space. The application of SOS1 in the proposed scheme consists of two selection levels: 
\begin{itemize}
    \item \textbf{Mode Selection:} A single SOS1 set is defined as \( \{s^{VV}, s^{VW}, s^{WV}\} \), ensuring only one mode is active at a time.
    \item \textbf{Segment Selection:} SOS1 sets are defined per mode:
    \begin{itemize}
        \item Volt-VAR: \( \{z_1^{VV}, \ldots, z_5^{VV}\} \),
        \item Volt-Watt: \( \{z_1^{VW}, z_2^{VW}, z_3^{VW}\} \),
        \item Watt-VAR: \( \{z_1^{WV}, \ldots, z_5^{WV}\} \).
    \end{itemize}
\end{itemize}
Mode and segment activations are linked via: \(  \sum z_l^{VV} = s^{VV}, \quad \sum z_l^{VW} = s^{VW}, \quad \sum z_l^{WV} = s^{WV}.    \)
\subsection{Real-Time Limited Measurement Implementation}

In practical applications, it is often unrealistic to assume full observability across all buses in a distribution system. Consequently, the LineDistOPF formulation presented in the previous section becomes inapplicable under such conditions. To address this limitation, we partition the system into two disjoint subsets: the \textit{observable set} and the \textit{unobservable set}. The observable set consists of buses where load, generation, and voltage measurements are available. The unobservable set includes buses for which direct measurement is not possible, where system states must be inferred or estimated. This classification enables us to restructure the LineDistOPF formulation for real-time implementation under limited measurement availability. The necessary condition to be applied here is that the controllable set must be a subset of the observable set—typically, the controllable set includes the substation and buses equipped with controllable DERs.

Beginning by partitioning all relevant vectors and matrices into their observable (superscript \(o\)) and unobservable (superscript \(u\)) components: \(Y=[Y^o \quad Y^u]^T, P^G=[P^{Go} \quad P^{Gu}]^T, Q^G=[Q^{Go} \quad Q^{Gu}]^T, A_0=[A_0^o \quad A_0^u]^T, A_1=[A_1^o \quad A_1^u]^T, A_2=[A_2^o \quad A_2^u]^T\). 
The matrices are similarly partitioned:
\begin{align*}
\mathbb{D}(P^L) &= 
\begin{bmatrix}
\mathbb{D}(P^{Lo}) & 0 \\
0 & \mathbb{D}(P^{Lu})
\end{bmatrix} \\
\mathbb{D}(Q^L) &= 
\begin{bmatrix}
\mathbb{D}(Q^{Lo}) & 0 \\
0 & \mathbb{D}(Q^{Lu})
\end{bmatrix}\\\quad
R^{eq} & =\begin{bmatrix}
    R^{oo} & R^{uo}\\R^{ou}&R^{uu}   
\end{bmatrix} , \quad X^{eq} = \begin{bmatrix}
    X^{oo} & X^{ou}\\X^{uo}&X^{uu}   
\end{bmatrix}
\end{align*}

Considering the original matrix (K) defined in Equation \ref{eq:FO3} Using the partitioned variables, the matrix \(K\) is reformulated as:
\begin{equation}
K = I + \begin{bmatrix}
    K^{oo}&K^{uo}\\K^{ou}&K^{uu}
\end{bmatrix}
\end{equation}

Where \(K^{oo},K^{ou},K^{uo},K^{uu}\) denote the contributions from the observable and unobservable powers to the observable to the unobservable voltages, respectively:
\begin{align}
\begin{bmatrix}
    K^{oo}\\ K^{ou}
\end{bmatrix}&= \begin{bmatrix}
    R^{oo}\\R^{ou}
\end{bmatrix} \, \mathbb{D} \left( \mathbb{D}(P^{Lo}) \left( A_1^o + \frac{A_2^o}{2\sqrt{Y_0^o}} \right) \right) \nonumber \\
&\quad + \begin{bmatrix}
    X^{oo}\\X^{ou}
\end{bmatrix} \, \mathbb{D} \left( \mathbb{D}(Q^{Lo}) \left( A_1^o + \frac{A_2^o}{2\sqrt{Y_0^o}} \right) \right) \label{eq:Ko}
\end{align}
\begin{align}
\begin{bmatrix}
    K^{uo}\\ K^{uu}
\end{bmatrix} &=\begin{bmatrix}
    R^{uo}\\R^{uu}
\end{bmatrix} \, \mathbb{D} \left( \mathbb{D}(P^{Lu}) \left( A_1^u + \frac{A_2^u}{2\sqrt{Y_0^u}} \right) \right) \nonumber \\
&\quad + \begin{bmatrix}
    X^{uo}\\X^{uu}
\end{bmatrix} \, \mathbb{D} \left( \mathbb{D}(Q^{Lu}) \left( A_1^u + \frac{A_2^u}{2\sqrt{Y_0^u}} \right) \right) \label{eq:Ku}
\end{align}

Now, we revisit the original linear equation used in the optimization model. By applying the partitioning of observable and unobservable variables, equation~\eqref{eq:FO3} can be reformulated as:
\begin{align}
Y^o = (I + K^{oo})^{-1} \left( R^{oo} P^o + X^{oo} Q^o + C_1\right)
\label{eq:FO3-partitioned}
\end{align}

Here, \(P^o\) and \(Q^o\) denote the net real and reactive power injections at the observable buses, including the substation and controllable DER buses. These net injections are computed as:
\(
P^{\text{o}}_i = P^G_i - P^L_i \left( a_0 + \frac{a_2}{2\sqrt{y_0}} \right)\), \(
Q^{\text{o}}_i = Q^G_i - Q^L_i \left( a_0 + \frac{a_2}{2\sqrt{y_0}} \right)
\) The residual term \(C_1\) in \eqref{eq:FO3-partitioned} accounts for the influence of unobservable loads and generators on the voltages at observable buses. It is defined as: \(
C_1 = R^{uo} P^u + X^{uo} Q^u - K^{uo} Y^u+Y_0
\) where \(P^u\) and \(Q^u\) are the net real and reactive power injections at unobservable buses. The unobservable voltage vector \(Y^u\) is itself a function of both observable and unobservable power injections, and is expressed as: \(
Y^u = (I + K^{uu})^{-1} ( R^{ou} P^o + R^{uu} P^u + X^{ou} Q^o + X^{uu} Q^u-K^{ou}Y^o +Y_0)
\).
Based on that, $C_1$ is indirectly influenced by the control variables because it includes terms dependent on the unobservable voltage vector \(Y^u\), which itself is a function of both observable and unobservable powers. Nevertheless, it is important to note that the matrix \(K^{uu}\), which appears in the expression for \(Y^u\), depends only on load values (i.e., fixed system parameters) and not on control variables. Therefore, during the relatively short computational time of the optimization process, it is reasonable to treat \(K^{uu}\) as constant. Given this observation, we can derive an updated expression for \(Y^o\) by substituting the functional form of \(Y^u\) into the residual term \(C_1\). The revised equation becomes:
\begin{align}
Y^o &= (I + K^{oo}-K_1 K^{ou})^{-1} [  \left(R^{oo} -K_1 R^{ou} \right) P^o  \notag \\
&\quad+ \left(X^{oo} -K_1 X^{ou} \right) Q^o + C_2 ]
\label{eq:FO3-partitioned2}
\end{align}

In this formulation, all variables are either measurable or known system parameters, except for \(C_2\) and \(K_{1}\). Where: 
\(C_2 = (I-K_1)Y_0 +(R^{uo}-K_1R^{uu})P^{u}+(X^{uo}-K_1X^{uu})Q^u\) and 
\(K_1=K^{uo}(1+K^{uu})^{-1}\) are independent on the control variables. To obtain a complete solution, these two quantities must be estimated first from the measurement and then used to optimize the control parameters. Noting that $K_1$ is a matrix and $C_2$ is a vector.

\subsection{Recursive Least Squares Estimation Using Algebraic Reformulation}

Recursive Least Squares (RLS) is a well-established technique for online parameter estimation in linear or linearized models. It is particularly well-suited for scenarios where measurements are sequentially available and computational efficiency is critical \cite{farhang2013adaptive}. In the context described by equation~\eqref{eq:FO3-partitioned2}, the parameters \(K_1\) (a matrix) and \(C_2\) (a vector) are not directly measurable and must be estimated from observable quantities \(Y^o\), \(P^o\), and \(Q^o\). 
Starting from the nonlinear model \ref{eq:FO3-partitioned2}, we left-multiply both sides by the matrix \( (I + K^{oo} - K_1 K^{ou}) \) and rearrange terms to yields to a linear model in the unknown parameters \( K_1 \) and \( C_2 \), and can be cast into the standard regression form:
\begin{equation}
\psi_t = \Phi_t \theta + \epsilon_t,
\label{eq:RLS_reformulated}
\end{equation}
where \( \psi_t \in \mathbb{R}^{n_o \times 1} \): the constructed output vector given by: \(
    \psi_t = R^{oo} P^o_t + X^{oo} Q^o_t - (I + K^{oo}) Y^o_t     \).  \( u_t \in \mathbb{R}^{n_u \times 1} \): the known input vector: \(
    u_t = R^{ou} P^o_t + X^{ou} Q^o_t - K^{ou} Y^o_t
    \).  \( \Phi_t \in \mathbb{R}^{n_o \times (n_o n_u + n_o)} \): the regressor matrix constructed from known system inputs:     \(
    \Phi_t = \begin{bmatrix}
    u_t^\top \otimes I_{n_o} & -I_{n_o}
    \end{bmatrix}
    \). \( \theta \in \mathbb{R}^{n_o n_u + n_o} \): the stacked parameter vector to be estimated:
    \(   \theta = [ \mathrm{vec}(K_1) \quad C_2]^T
    \). \( \epsilon_t \in \mathbb{R}^{n_o \times 1} \): the residual error at time step \( t \), modeled as zero-mean white noise. The RLS algorithm is then applied to update the parameter estimate \( \hat{\theta}_t \) at each time step as follows:
\begin{subequations}
    \begin{align}
    K_t &= P_{t-1} \Phi_t^\top \left( \lambda I + \Phi_t P_{t-1} \Phi_t^\top \right)^{-1} \\
    \hat{\theta}_t &= \hat{\theta}_{t-1} + K_t \left( \psi_t - \Phi_t \hat{\theta}_{t-1} \right) \\
    P_t &= \lambda^{-1} \left( P_{t-1} - K_t \Phi_t P_{t-1} \right)
    \end{align}
    \label{eq:RLS_update}
\end{subequations}
 where: \( \hat{\theta}_t \): updated estimate of the parameter vector. \( P_t \): error covariance matrix representing estimation confidence.  \( \lambda \in (0, 1] \): forgetting factor that emphasizes recent data. The matrix \( K_t \) acts as the adaptive gain determining how much the estimate is adjusted at each time step based on new information. Once \( \hat{\theta}_t \) is obtained, the estimates of \( K_1 \) and \( C_2 \) can be recovered by: \(
\hat{K}_1 = \mathrm{reshape}(\hat{\theta}_{1:n_o \cdot n_u}, n_o, n_u), \quad \hat{C}_2 = \hat{\theta}_{n_o \cdot n_u + 1 : \, \text{end}}
\). 
From a mathematical perspective, System losses (which are ignored in the LineDistOPF formulation) can be represented by incorporating a nonlinear loss function dependent on the control variables (e.g., \(P^o\), \(Q^o\)) into equation~\eqref{eq:FO3-partitioned2}. The resulting expression can then be cast into a regression structure, as presented in Equation~\eqref{eq:RLS_reformulated}, allowing for implicit modeling of nonlinear behaviors through the RLS estimation framework.

\section{T\&D Framework}

\begin{figure}
    \centering
    \includegraphics[width=\linewidth]{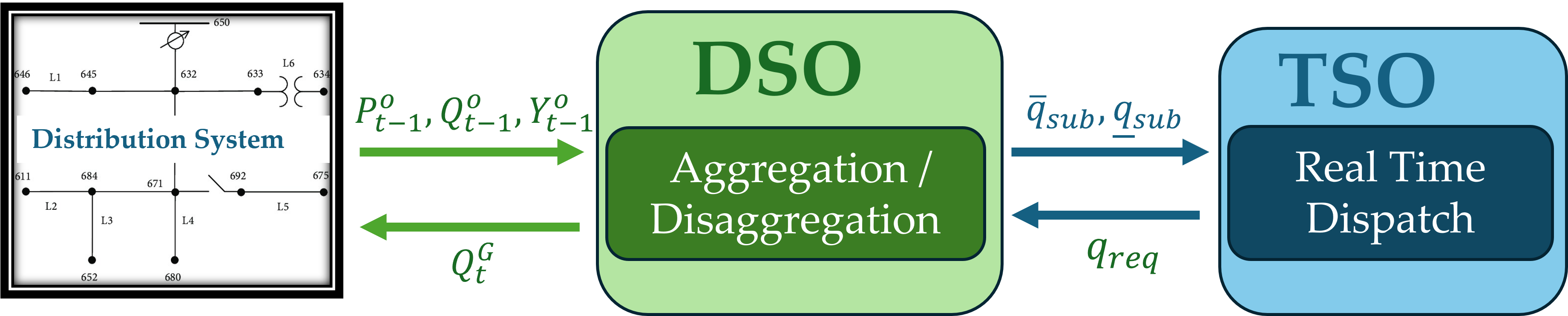}
    \caption{Three-step T\&D framework }
    \label{fig:TD}
\end{figure}

The proposed {Transmission--Distribution (T\&D)} coordination dispatch algorithm operations through three functions, shown in Fig.~\ref{fig:TD}: \emph{DSO Aggregation/Disaggregation}, and \emph{TSO dispatch}. Its goal is to determine the optimal dispatch, control mode, and droop parameters of PV inverters under limited observability.
To minimize real power curtailment while maximizing voltage support, a hierarchical MILP structure is used. Firstly, the maximum DER generation is optimized in stage 1, then the Aggregation and disaggregation are found in stage 2 using the following models. The control variables ($\mathbb{C}$) of the following models are explained in Section III. 

\textbf{Stage 1: Real Power Maximization}
\begin{subequations}
    \begin{align}
      P^* = \max_{\mathbb{C}} \quad & \sum_i P_i  \label{eq1a:objective}
        \\ \text{subject to} \quad & (\ref{eq:cap}- \ref{eq:wv}, \ref{eq:MILP}, \ref{eq:FO3-partitioned2}) \label{eq1a:cons1}\\& 
        \underline{V}^2 \leq Y^o \leq \overline{V}^2 \label{eq1a:cons2}\\& 
        P_i \leq P_{\text{max},i}\label{eq1a:cons3}
         \end{align}
         \label{eq:stage1}
\end{subequations}

Where: \ref{eq1a:cons1} grantees the local controller feasibility and distribution network power flow constraints, \ref{eq1a:cons1} forced distribution voltages to be within the normal range of voltage $[\underline{V}, \overline{V}]$ and \ref{eq1a:cons3} to make sure that the DER output is less than the maximum real power based on the available maximum power $(P_{max})$. 

\noindent\textbf{Stage 2a: DSO aggregation}
\begin{subequations}
    \begin{align}
        [\underline{Q}, \overline{Q}] = \min_{\mathbb{C}}/\max_{\mathbb{C}} \quad & \sum Q^{\text{G}}_i  \label{eq1b:objective}
        \\ \text{subject to} \quad & (\ref{eq1a:cons1}- \ref{eq1a:cons3}) \label{eq1b:cons1}\\& 
          P^* = \sum_i P_i  \label{eq1b:cons2}
         \end{align}
         \label{eq:Aggregation}
\end{subequations}

The same constraints from Stage 1 are applied in \ref{eq1b:cons1} and \ref{eq1b:cons2} to ensure that the maximum real power is maintained. The outcome of Stage 2a determines the aggregated distribution-level reactive power limits, $[\underline{Q}, \overline{Q}]$, which characterize the system’s flexibility at the TSO interface. Based on each distribution network’s capability curve, the TSO performs real-time dispatch to optimize the requested reactive power from each distribution network by solving:

\noindent\textbf{ TSO Real Time Dispatch}
\begin{subequations}
    \begin{align}
         &\min_{Q^L,Q^G} \quad   \sum_{k \in \mathcal{BI}} c_v (V_{k} - V_{\text{setpoint}})^2+c_q (Q_{k} - 0)^2  \label{eq2:outer_objective} & \\
         &\text{subject to}  \notag \\
         & g(\mathcal{X}) = 0, \quad h(\mathcal{X}) \leq 0, \label{eq2:power_flow_constraints} & \\
    & \underline{Q}^L_{i} \leq Q^L_i\leq \overline{Q}^L_i, \quad  \forall i \label{eq2:qk_limits1}
    \end{align}
    \label{eq:Dispatch}
\end{subequations}
Where $c_v$ and $c_q$ are sensitivity weights and $\mathcal{BI}$ refers to the set of load and generator buses. This optimization problem aims to minimize voltage deviations across the network by adjusting the reactive power from loads \(Q^L\) and generators \(Q^G\) with minimum reactive power. Constraint \eqref{eq2:power_flow_constraints} encompasses power flow equations and operational constraints related to branch flows, voltage levels, and generator limits. Constraint \eqref{eq2:qk_limits1} ensures the reactive power at each distribution network stays within the capability defined in stage 2a, represented as \(\underline{Q}^L_i\) and \(\overline{Q}^L_i\). Minimizing voltage deviations across the network and reactive power is commonly used in the literature as an objective function in traditional TSO dispatch algorithms \cite{liu2023time}.

\noindent\textbf{Stage 2b: DSO disaggregation}
\begin{subequations}
    \begin{align}
        \min_{\mathbb{C}} \quad & \sum_i w_i |Q^G_{i} | \label{eq1c:objective}
        \\ \text{subject to} \quad & (\ref{eq1b:cons1}, \ref{eq1b:cons2}) \label{eq1c:cons1}\\& 
          Q_{\text{sub}} = Q_{\text{req}}  \label{eq1c:cons2}\\&
          w_i = 1 - \frac{\partial Q_{\text{sub}}/\partial q_i}{\sum_j (\partial Q_{\text{sub}}/\partial q_j)}
         \end{align}
         \label{eq:diaggregation}
\end{subequations}

Here, the same constraints from Stage 2a are implemented in \ref{eq1c:cons1} and \ref{eq1c:cons2} to maintain the requested reactive power from the TSO, obtained through the TSO dispatch step at the substation. The weight $w_i$ represents each DER’s sensitivity to variations in substation reactive power, prioritizing support from units located closer to the substation.

\subsection{Iterative Coordination Algorithm}
Fig.~\ref{fig:flowchart} summarizes the iterative coordination between the DSO and TSO. The substation acts as the interface for data exchange and control actions.


\tikzstyle{block1} = [rectangle, draw, fill=blue!10, text width=5cm, text centered, rounded corners, minimum height=2em]
\tikzstyle{block3} = [rectangle, draw, fill=white, text width=5cm, text centered, rounded corners, minimum height=2em]
\tikzstyle{decision} = [diamond, draw, fill=orange!10, text width=3cm, text centered, minimum height=2em, aspect=2]
\tikzstyle{line} = [draw, -latex']
\tikzstyle{cloud} = [draw, ellipse,fill=green!10, node distance=2cm, minimum height=2em]
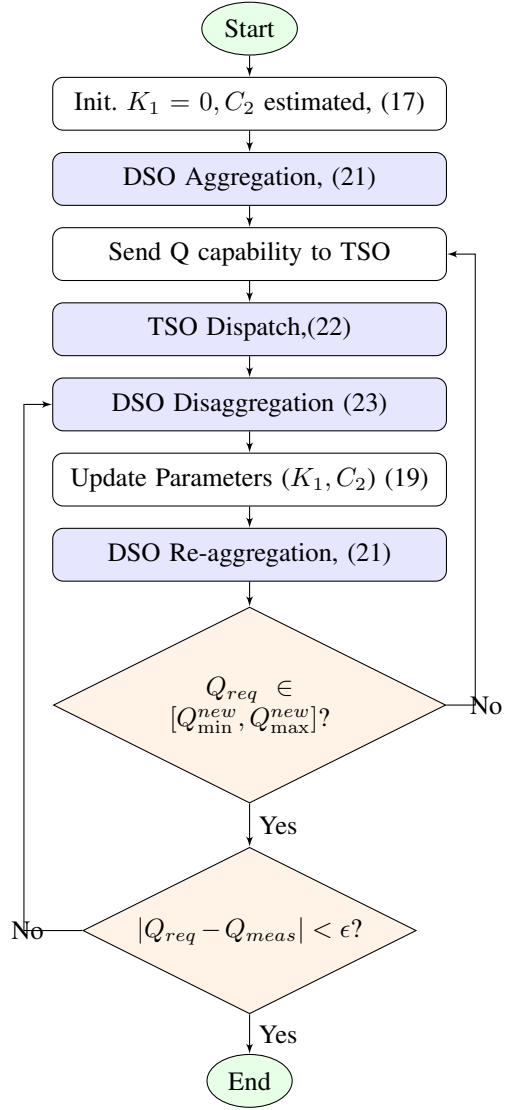
\begin{figure}
    \centering
\begin{tikzpicture}[node distance = 1cm, auto]
    \node [cloud] (start) {Start};
    \node [block3, below of=start] (init) {Init. $K_1=0, C_2$ estimated, (\ref{eq:FO3-partitioned2})};
    \node [block1, below of=init] (dso) {DSO Aggregation, (\ref{eq:Aggregation})};
    \node [block3, below of=dso] (sendQ) {Send Q capability to TSO};
    \node [block1, below of=sendQ] (tso) {TSO Dispatch,(\ref{eq:Dispatch})};
    \node [block1, below of=tso] (disagg) {DSO Disaggregation (\ref{eq:diaggregation})};
    \node [block3, below of=disagg] (update) {Update Parameters ($K_1, C_2$) (\ref{eq:RLS_update})};
    \node [block1, below of=update] (dso2) {DSO Re-aggregation, (\ref{eq:Aggregation})};
    \node [decision, below of=dso2, yshift=-1cm] (decision) { $Q_{req} \in [Q_{\min}^{new}, Q_{\max}^{new}]$?};
    \node [decision, below of=decision, yshift=-2cm] (decision2) { $|Q_{req}-Q_{meas}|<\epsilon$?};
    \node [cloud, below of=decision2, yshift=-0cm] (end) {End};

    \path [line] (start) -- (init);
    \path [line] (init) -- (dso);
    \path [line] (dso) -- (sendQ);
    \path [line] (sendQ) -- (tso);
    \path [line] (tso) -- (disagg);
    \path [line] (disagg) -- (update);
    \path [line] (update) -- (dso2);
    \path [line] (dso2) -- (decision);
    \path [line] (decision) -| node[near start, right] {No} ++(3,0) |- (sendQ);
    \path [line] (decision) -- node[right] {Yes} (decision2);
    \path [line] (decision2) -| node[near start, left] {No} ++(-3,0) |- (disagg);
    \path [line] (decision2) -- node[right] {Yes} (end);
\end{tikzpicture}
    \caption{Flowchart of the enhanced TDCDA.}
    \label{fig:flowchart}
\end{figure}

\begin{enumerate}
    \item \textbf{Initialization:} The DSO initializes parameters ($K_1=0$, $C_2$ estimated) using (\ref{eq:FO3-partitioned2}) based on current measurements.
    \item \textbf{Aggregation:} The DSO solves (\ref{eq:Aggregation}) to obtain $[\underline{Q}, \overline{Q}]$ and communicates it to the TSO.
    \item \textbf{Dispatch:} The TSO solves (\ref{eq:Dispatch}) to determine $Q_{\text{req}}$ and sends it to the DSO.
    \item \textbf{Disaggregation:} The DSO executes (\ref{eq:diaggregation}) to allocate $Q_{\text{req}}$ among DERs or loads.
    \item \textbf{Update and Re-aggregation:} Parameters ($K_1$, $C_2$) are updated via RLS (\ref{eq:RLS_update}); new capabilities are recalculated.
    \item \textbf{Convergence Check:} The process repeats until $Q_{\text{req}} \in [Q_{\min}^{new}, Q_{\max}^{new}]$ and $|Q_{\text{req}} - Q_{\text{meas}}| < \epsilon$.
\end{enumerate}

\section{SIMULATION RESULTS AND ANALYSIS}



 To demonstrate and validate the proposed algorithm, a different set of simulation scenarios has been established. For the distribution network, two standard IEEE test systems are selected: IEEE 13-Bus Test System: The system includes 9 single-phase smart inverter-based DERs, each rated at 300~kW, placed at buses 634, 675, and 680, and IEEE 123-Bus Test System: 45 single-phase DERs were installed at locations specified in \cite{Ref8}. 
For the transmission network, the IEEE 9-Bus test system is chosen to enable a clear examination of the DSO-TSO interaction. The test system load configuration of 90 MW at bus 5, 100 MW at bus 7, and 125 MW at bus 9. Distribution systems interfacing with this transmission network are integrated at buses 5, 7, and 9, comprising 29, 32, and 40 instances of the IEEE 13-bus/IEEE 123-bus distribution networks, respectively. All simulations are conducted using an integrated platform where distribution-level simulation is handled by OpenDSS coupled with Python, and the optimization problem is solved using the Gurobi optimization engine in Python. Transmission-level power flow simulations leverage the PandaPower toolbox interfaced through PowerModels in Julia, connected via a Python-Julia bridge. 

\subsection{Real-Time Implementation Feasibility}

The computational efficiency and feasibility of real-time implementation are assessed by measuring solver runtimes across various optimization steps (real power maximization $P^*$, stage 2a\_min, stage 2a\_max, stage 2b). Results demonstrate the significant improvements achieved by applying SOS1 compared to the Binary method, especially for large systems like IEEE 123. Table~\ref{tab:runtime} summarizes the runtime and iterations of the solver. For the IEEE 13-bus system, both SOS and Binary formulations demonstrate very fast runtimes and low iteration counts, suitable for real-time implementation. However, for the IEEE 123-bus system, the SOS formulation leads to significantly lower iteration counts and runtime compared to the binary formulation, suggesting that SOS is more suitable for real-time dispatch applications in large-scale distribution systems.

\begin{table}[htbp]
\centering
\caption{solver runtime and iterations for test systems}
\label{tab:runtime}
\scriptsize
\begin{tabular}{|l|l|c|c|}
\hline
\textbf{System} & \textbf{Stage} & \textbf{Runtime (s)} & \textbf{Iterations} \\ \hline
\multirow{4}{*}{IEEE 13-bus (SOS)} 
 & $P^*$       & 0.071 & 1   \\ \cline{2-4}
 & stage 2a\_min & 0.071 &37   \\ \cline{2-4}
 & stage 2a\_max & 0.068 & 1  \\ \cline{2-4}
 & stage 2b      & 0.0079 & 12  \\ \hline

\multirow{4}{*}{IEEE 13-bus (Binary)} 
 & $P^*$       & 0.081 & 1    \\ \cline{2-4}
 & stage 2a\_min & 0.076 & 355  \\ \cline{2-4}
 & stage 2a\_max & 0.065 & 1  \\ \cline{2-4}
 & stage 2b      & 0.0123 & 31   \\ \hline

\multirow{4}{*}{IEEE 123-bus (SOS)} 
 & $P^*$       & 0.202& 1  \\ \cline{2-4}
 & stage 2a\_min & 0.211 & 190   \\ \cline{2-4}
 & stage 2a\_max & 0.209 & 178   \\ \cline{2-4}
 & stage 2b      & 0.051 & 61   \\ \hline

\multirow{4}{*}{IEEE 123-bus (Binary)} 
 & $P^*$       & 0.322 & 1   \\ \cline{2-4}
 & stage 2a\_min & 1.956 & 21545   \\ \cline{2-4}
 & stage 2a\_max & 0.308 & 1530  \\ \cline{2-4}
 & stage 2b      & 0.0872 & 386  \\ \hline

\end{tabular}
\end{table}

\subsection{Effectiveness of Droop Control Integration}

\textcolor{black}{ Table III compares the maximum active and reactive power capability obtained under five operating scenarios: the three individual IEEE 1547 default droop modes (Volt–Var, Volt–Watt, and Watt–Var), an uncoordinated free P–Q operating mode, and the optimized droop-based operation. The first three modes apply the standard IEEE 1547 curves without any tuning, meaning that each inverter follows a fixed voltage-dependent or power-dependent characteristic with no adaptation to system needs; the only optimization is the set point. In contrast, the free P–Q mode assumes that the inverter can fully utilize its theoretical capability curve without considering the internal behavior of local controllers. Finally, the optimized mode jointly selects the most effective droop configuration and tunes its parameters to maximize flexibility while ensuring feasibility.}

\textcolor{black}{ In terms of real power delivery, the optimized and free P–Q cases both reach the maximum available real power of 1.98 MW per inverter, representing the upper limit at the assumed penetration level. This demonstrates that, in the absence of restrictive control curves, DERs can fully utilize their power potential. However, under the default Volt–Var mode, the maximum real power is reduced to 1.7 MW, resulting in a curtailment of more than 14\%. A similar, although smaller, reduction occurs under Volt–Watt mode, which caps real power at 1.8 MW. These reductions highlight the operational inefficiency introduced when DERs are forced to follow fixed IEEE 1547 droop curves without coordination or optimization.}

\textcolor{black}{ From a reactive capability perspective, the results reveal an even sharper contrast. The optimized setting achieves a reactive span from 1.4 MVAR injection down to –0.9 MVAR absorption, providing roughly 2.3 MVAR of usable reactive flexibility. The free P–Q mode shows bigger limits because it assumes the full capability curve is accessible. In comparison, the Volt–Var and Volt–Watt modes offer significantly narrower ranges, with Q${\max}$ limited to 0.7 MVAR and 1.1 MVAR, respectively, and Q${\min}$ bounded at 0 or –0.5 MVAR. Under Watt–Var mode, reactive capability collapses to a fixed value of –0.9 MVAR, eliminating any voltage support flexibility. This collapse occurs because the reactive power capability becomes tightly constrained when real power is pushed near its limit.}

\textcolor{black}{ These observations underscore that the freely assumed P–Q operation, although mathematically appealing for optimization, produces unrealistic expectations because the local inverter controller cannot physically track arbitrary setpoints that violate its embedded local controller. In contrast, when the droop mode or curve parameters are optimized, the inverter preserves maximum feasible flexibility while ensuring that all dispatched setpoints remain dynamically achievable. These results show that achieving maximum flexibility requires at least one of the following actions: (i) optimizing the droop curve parameters or (ii) appropriately selecting the operating mode from the standard IEEE 1547 default curves. Furthermore, the analysis indicates that any two control modes among VV, VW, and WV are sufficient to realize the maximum achievable flexibility, meaning that a full set of all three modes is not necessary. We also conducted a similar analysis on the IEEE 123-bus system. In that case, the default VV mode failed to bring voltages back into the acceptable range, indicating that voltage violations persist even with curtailment when DERs are not adaptively controlled and optimized. This highlights the need to optimize DER control parameters rather than relying on static settings. The only droop curve capable of bringing voltages within the acceptable range in the IEEE-123 bus test system is the Watt-VAR control, and also real power curtailment is necessary. }

\begin{table}[]
\centering
\label{table:Comparisne}
\caption{Comparison of Droop Control Modes, Free PQ Operation, and Optimized Droop Settings}
\begin{tabular}{lccc}
\hline
\textbf{Control Mode} & \textbf{P$_{\max}$ (MW)} & \textbf{Q$_{\max}$ (MVAR)} & \textbf{Q$_{\min}$ (MVAR)}  \\ 
\hline
Volt--Var   & 1.7 & 0.7  &   0    \\ 
Volt--Watt  & 1.8 & 1.1  & -0.5   \\ 
Watt--Var    & 1.9 & -0.9 & -0.9  \\ 
P--Q Free       & 1.98 & 1.4  & -1.0  \\ 
Optimized           & 1.98 & 1.4  & -0.9    \\ 
\hline
\end{tabular}
\label{tab:droop_comparison}
\end{table}

\subsection{Evaluation of Coordination with TSO}

\begin{figure}
    \centering
    \includegraphics[width=0.8\linewidth]{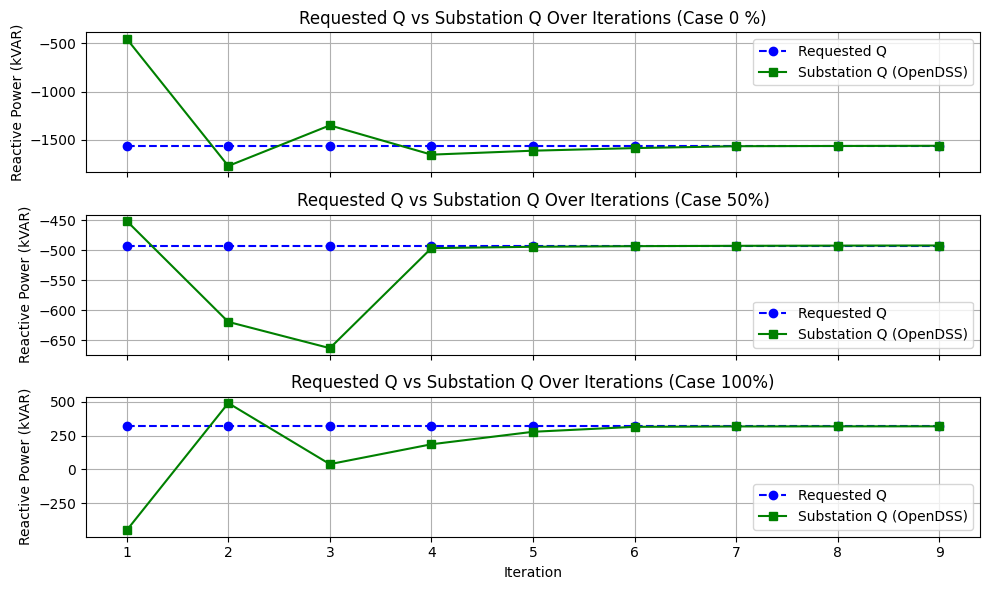}
    \caption{Iterative convergence behavior between TSO \& DSO coordination.}
    \label{fig:iterative_convergence}
\end{figure}

To illustrate the effectiveness of the iterative coordination process, the convergence behavior of the reactive power dispatch between the TSO and DSO is analyzed. Figure~\ref{fig:iterative_convergence} shows the reactive power at the substation and the requested values for three scenarios: when the TSO requests 0\% of the capability curve (the minimum value), 50\% (a midpoint), and 100\% (the maximum value of the capability curve), displayed in the lower, middle, and upper subplots, respectively.  These results correspond to the IEEE 13-bus test system and demonstrate that the dispatched reactive power closely tracks the TSO request with a small error after only 5 DSO iterations. A similar analysis is performed for the IEEE 123-bus system, where the algorithm required 10 iterations to achieve comparable accuracy. The figure also clearly shows that losses---being a nonlinear term---are indirectly accounted for through the iterative estimation of \(C_2\) and \(K_1\).

\begin{figure}
    \centering
    \includegraphics[width=0.4\textwidth]{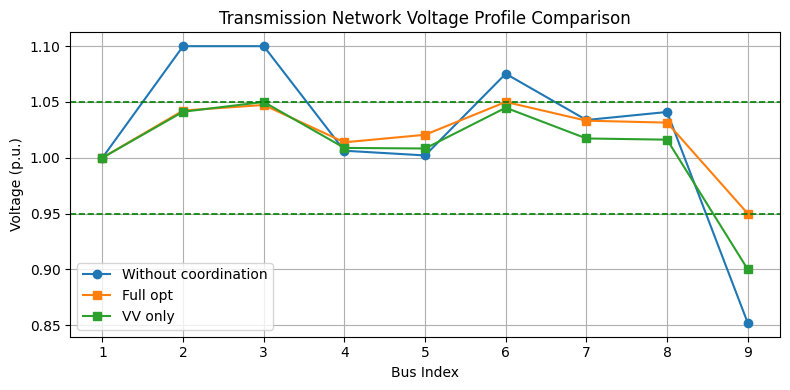}
    \caption{Voltage profile of the 9-bus transmission system under line 4--9 outage for different coordination and control scenarios at 20\% PV.}
    \label{fig:voltage_profile_coordination}
\end{figure}

To see the impact of local controllers on the transmission system, Figure~\ref{fig:voltage_profile_coordination} shows the voltage profile of the 9-bus transmission system following the outage of line 4--9 under three scenarios.  
Without TSO--DSO coordination, several buses experience voltage violations beyond the acceptable range (0.95--1.05~p.u.).  
When full optimization is applied—coordinating multiple control modes—the voltage profile remains within limits across all buses, confirming effective reactive power support.  
The Volt--Var-only case partially improves the profile but still approaches the upper and lower limits, highlighting that coordinated multi-mode control provides the most stable voltage response under contingencies.

\subsection{Scalability Analysis}

In this subsection, the scalability of the proposed algorithm is evaluated by comparing its computational performance and resource utilization across different system sizes. Specifically, the number of single-phase DERs is varied from 3 to 168 units in the IEEE-123 test system and from 3 to 21 units in the IEEE-13 bus test system to evaluate how the solver's performance scales with increasing DER numbers at different systems.

\begin{figure}[ht]
    \centering
    \includegraphics[width=0.8\linewidth]{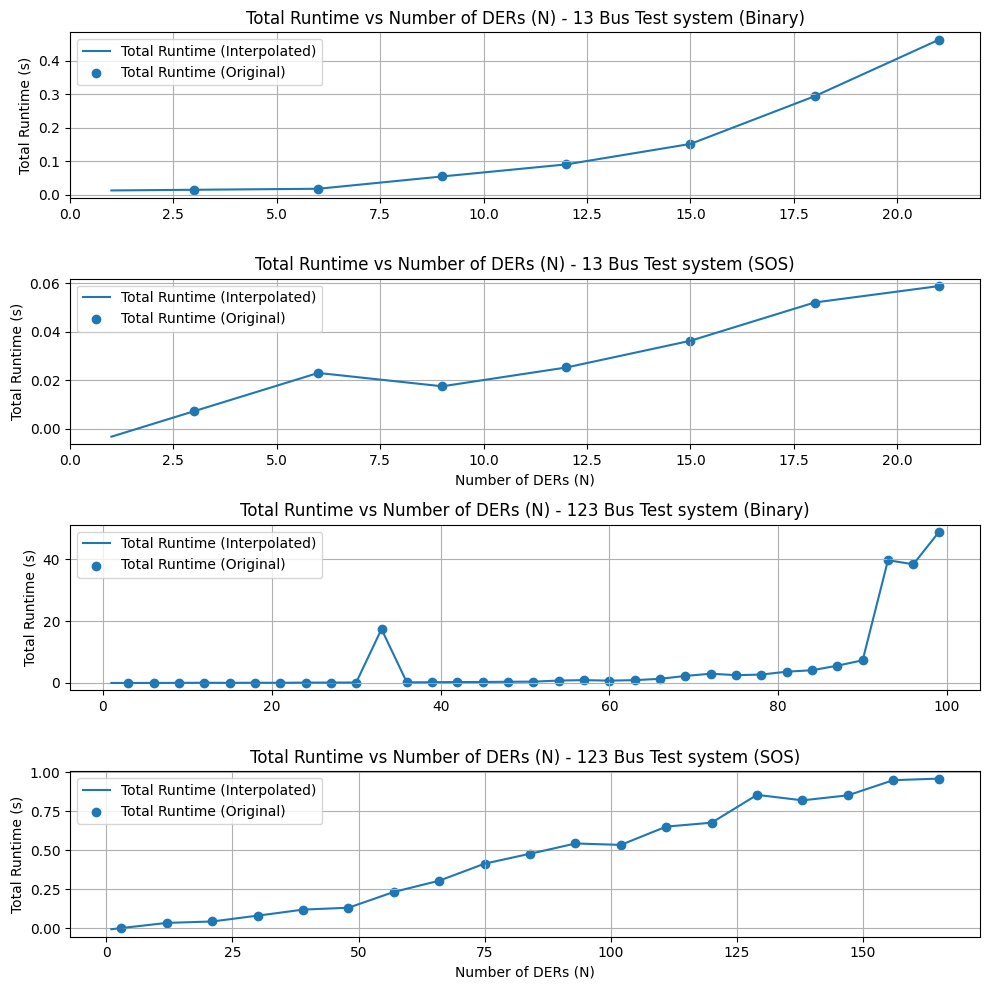}
    \caption{Computational runtime for IEEE-13 and IEEE 123-bus using Binary ans SOS methods.}
    \label{fig:scalability_runtime}
\end{figure}

To check on the computational performance, Figure~\ref{fig:scalability_runtime} presents the total computational runtime across all optimization stages for both test systems, comparing the performance of the binary MILP model and the SOS-based formulation. The results show that the computational time for the binary MILP model increases exponentially with the number of DERs, whereas the SOS-based model exhibits a linear growth trend. In the IEEE 123-bus test system, the binary MILP model becomes intractable when attempting to solve for 105 single-phase DERs, it was run for three hours without reaching an optimal solution. Where as the optimization of 159 DER using SOS method run with less than a second.

To further understand the algorithm's computational requirements, Table~\ref{tab:memory_usage} provides typical data on memory usage and computational resource demand for each optimization scenario, comparing results across IEEE 13-bus and IEEE 123-bus systems.
Table \ref{tab:memory_usage} presents a comparative analysis of CPU time and memory usage for the proposed reactive power dispatch algorithm under both SOS-based and binary MILP formulations. For the IEEE 13-bus system, both formulations exhibit low CPU times and modest memory usage, with SOS slightly increasing memory demand as the number of DERs grows but maintaining better scalability in runtime. In contrast, for the IEEE 123-bus system, the SOS formulation demonstrates significantly better computational efficiency as the number of DERs increases. While memory usage under SOS grows with the system size, it remains stable beyond 81 DERs, suggesting an efficient solver implementation. The binary formulation, on the other hand, becomes increasingly computationally intensive, with CPU time rising sharply from 27.17 seconds for 81 DERs to 87.57 seconds for 96 DERs, indicating scalability issues. These results confirm the advantage of the SOS-based method in achieving real-time feasibility for large-scale systems, even at the cost of slightly higher memory consumption.

\begin{table}
\centering
\caption{Memory and computational resource usage comparison}
\label{tab:memory_usage}
\scriptsize
\begin{tabular}{|l|c|c|c|}
\hline
\textbf{System} & \textbf{Number of DER} & \textbf{CPU Time (s)} & \textbf{ Memory (MB)} \\ \hline

\multirow{4}{*}{IEEE 13-bus (SOS)} 
 & 3      & 1.1& 2053.29   \\ \cline{2-4}
 & 9 & 1.49 &2053.68  \\ \cline{2-4}
 & 15& 2.05& 2054.09  \\ \cline{2-4}
 & 21     & 2.27 &2058.25 \\ \hline

\multirow{4}{*}{IEEE 13-bus (Binary)} 
 & 3      & 0.98& 1786.79    \\ \cline{2-4}
 & 9 & 1.4 & 1841.25  \\ \cline{2-4}
 &15&2.02 & 1935.44  \\ \cline{2-4}
 & 21      & 3.68 & 2053.18   \\ \hline

\multirow{4}{*}{IEEE 123-bus (SOS)} 
 & 42      & 6.94 & 3557.99    \\ \cline{2-4}
 & 81 & 23.1 & 6739.07   \\ \cline{2-4}
 & 120 & 38.33 & 6731.91   \\ \cline{2-4}
 & 169    & 38.3& 6709.93  \\ \hline

\multirow{4}{*}{IEEE 123-bus (Binary)} 
 & 42      & 6.93 & 2438.86  \\ \cline{2-4}
 &81 & 27.17 & 5249.12   \\ \cline{2-4}
 & 90 & 37.2 & 5554.25 \\ \cline{2-4}
 & 96     & 87.57 & 5926.75 \\ \hline

\end{tabular}
\end{table}

\section{Conclusion}
This paper presented the real-time TSO–DSO coordination framework for reactive power dispatch in systems with DER penetration. The proposed method integrates IEEE 1547-compliant droop control modes of smart inverters—Volt-VAR, Volt-Watt, and Watt-VAR—into a sensitivity-aware MILP formulation enhanced with Special Ordered Sets (SOS1) for scalable, real-time operation. The framework incorporates Recursive Least Squares (RLS) estimation to accommodate limited observability, making it suitable for practical deployment in distribution systems. Simulation studies on IEEE 13-bus and 123-bus networks demonstrate that the approach enables flexible, low-curtailment reactive power dispatch and improves voltage regulation across the TSO–DSO interface.

\bibliographystyle{IEEEtran}
\bibliography{ref_jurnal}

\end{document}